\documentclass[pra,aps,twocolumn,superscriptaddress,showpacs,showkeys,nofootinbib]{revtex4-1}
\usepackage{amsmath,amsthm,amssymb,graphicx,subfigure,xcolor}
\usepackage{graphics,float}
\usepackage[unicode=true]{hyperref}
\usepackage{booktabs}
\usepackage{tabularx}
\usepackage{color}
\usepackage{tabu}
\usepackage{multirow}
\usepackage{graphicx}
\usepackage{subfigure}
\usepackage{setspace}
\usepackage{amsmath}
\makeatletter

\vspace{-0.8cm}
\newtheorem{theorem}{\indent Theorem}
\newtheorem*{theorem*}{\indent Theorem}

\newtheorem*{corollary*}{\indent Corollary}
\newtheorem{lemma}{\indent Lemma}
\newtheorem*{lemma*}{\indent Lemma}
\newtheorem{proposition}{\indent Proposition}
\newtheorem*{proposition*}{\indent Proposition}
\theoremstyle{definition}
\newtheorem{definition}{\indent Definition}
\newtheorem*{definition*}{\indent Definition}
\theoremstyle{remark}

\newtheorem*{remark*}{\indent Remark}

\begin{document}

\title{The block-coherence measures and the coherence measures based on  positive-operator-valued measures }

\author{Liangxue Fu}
\author{Fengli Yan}
\email{flyan@hebtu.edu.cn}
\affiliation {College of Physics, Hebei Key Laboratory of Photophysics Research and Application, Hebei
Normal University, Shijiazhuang 050024, China}
\author{Ting Gao}
\email{gaoting@hebtu.edu.cn}
\affiliation {School of Mathematical Sciences, Hebei
Normal University, Shijiazhuang 050024, China}

\begin{abstract}
We mainly study the block-coherence measures based on resource theory of block-coherence  and the coherence measures based  on positive-operator-valued measures (POVM). Several block-coherence measures including a block-coherence measure  based on maximum relative entropy, the one-shot block coherence cost under the maximally block-incoherent operations, and a  coherence measure based on coherent rank have been introduced and  the relationships between these block-coherence measures have been obtained. We also give the definition of the maximally block-coherent state and describe the deterministic coherence dilution process by constructing block-incoherent  operations.
Based on the  POVM coherence resource theory, we propose a POVM-based coherence measure by using the known scheme of building POVM-based coherence measures from block-coherence measures, and  the one-shot block coherence cost under the maximally POVM-incoherent operations.  The relationship between the POVM-based  coherence measure and the  one-shot block coherence cost under the maximally POVM-incoherent operations is analysed.
\end{abstract}



\maketitle

\section{INTRODUCTION}

Quantum coherence is an important ingredient in quantum information processing \cite{1}. Baumgratz $\emph{et~al.}$ proposed the theoretical framework of the resource theory of quantum coherence in 2014 (BCP framework) \cite{2}. The theoretical framework comprises three basic elements: a set of free states which do not contain resource, a corresponding set of free operations that map an arbitrary  free state to a free state (generating no resource), and a metric functional \cite{2}.

In the resource theory of quantum coherence, the free states are the incoherent states, which  can be diagonalized under a fixed reference basis \cite{2}. The free operations (incoherent operations) are some specified classes of physically realizable operations \cite{2}. According to different operational capabilities and physical relevance, the sets of free operations may be : the maximally incoherent operation  \cite{3,4}, the dephasing-covariant incoherent operation \cite{3,5,6}, the incoherent operation  \cite{2}, the strictly incoherent operation  \cite{7,8}, and the physically implementable incoherent operation  \cite{6}. In order to quantify coherence, different coherence measures are defined in the resource theory of coherence, such as $l_{1}$-norm coherence measure \cite{2}, relative entropy coherence measure \cite{2} and coherence of formation \cite{7,9}, etc. Coherence measures of different meanings help us better quantify and understand coherence \cite{2,3,4,5,6,7,8,9,10,11,12, ZhangGaoYan}.

An interesting problem in the resource theory of quantum coherence is the transformation of states via free operations, specially the transformation between an arbitrary state $\rho$ and a maximally coherent  state \cite{13}.  In particular, the process of converting a given state $\rho$ to the maximally coherent state by incoherent operation is referred to as coherence distillation \cite{13,14,15,16}. In contrast to distillation, the dilution process converts the maximally coherent state into the desired target state \cite{13,17,18}. The processes of asymptotic dilution and distillation are performed under the independent and identically distributed  assumption \cite{13,15,16,17}, which ignores the possible correlation between different state preparations. Therefore, in order to relax the assumption, it is necessary to consider the one-shot scenario, where only one copy of the state is supplied \cite{13,15,16,17}.

The resource theory of block-coherence  was introduced in Ref. \cite{4}. Here we adopt the framework proposed in Ref. \cite{19}. In the resource theory of block-coherence, the block-incoherent states can be considered to be generated by a von Neumann measurement $\mathbf{P}=\{P_{i}\}$, $i=1,2,\cdots,d$,  i.e., the block-incoherent state $\sigma=\sum_{i=1}^{d}P_{i}\rho P_{i}$ for state $\rho\in \mathcal{S}$, where $\mathcal{S}$ denotes the set of quantum states on the Hilbert space $\mathcal{H}$,  the rank of the orthogonal projector $P_{i}$ is arbitrary and the orthogonal projectors form a complete set, i.e., $\sum_{i=1}^{d}P_{i}=\mathbb{I}$ \cite{4,19,20,21}.

 In 2019, Bischof $\textit{et al}$. \cite{19} established the resource theory of coherence based on the positive-operator-valued measures (POVM). The theory is called POVM coherence resource theory. The approach of this theory is to employ the Naimark extension to define the POVM coherence via the  block-coherence in a larger Hilbert space, where the quantum states act through an embedded channel in the $d'$-dimensional ($d' > d$) Hilbert space $\mathcal{H}'$ (Naimark space), and a POVM $\mathbf{E}$ is extended to the projection measurement $\mathbf{P}$ of the Naimark space $\mathcal{H}'$ \cite{19,20,22}. We will give detailed description of the resource theory of block-coherence  and the POVM coherence resource theory in the second section of the article.

In this paper, we mainly study the block-coherence measures  based on resource theory of block-coherence  and the coherence measures  based on POVM coherence resource theory and analyse  the relationship between these block-coherence measures.

The paper is divided into five sections. In Section II, we give  main concepts, review the resource theory of block-coherence  and POVM coherence resource theory. In Section III,   we define two block-coherence measures and the one-shot block coherence cost  in the framework of resource theory of block-coherence, and analyse the relationship between  these coherence measures. We illustrate the problem of deterministic coherence dilution by constructing block-incoherent operation. In Section IV,  a POVM-based coherence measure  and  the  one-shot block coherence cost under the maximally POVM-incoherent operations are defined and analysed.

\section{Background}\label{sec:review}

\subsection{Block-coherence theoretical framework}\label{sec:review}

In 2006, {\AA}berg introduced the general measurement method of superposition degree of mixed quantum states and applied it to the orthogonal decomposition of Hilbert space, created the resource theory of block-coherence. Similar to the theoretical framework of BCP, it also consists of  three elements: the set of block-incoherent states,  the set of block-incoherent operations, and the block-coherence measures \cite{4,19}.

 The Hilbert space $\mathcal{H}$ is divided into $d$ orthogonal subspaces, the projective measurement $\mathbf{P}=\{P_{i}\}$ is performed on a set $\mathcal{S}$ of quantum states, where $P_{i}$ is the projector of the $i$th subspace. Block-incoherent states \cite{4,19,20,21} are defined as
\begin{equation}\label{7}
\begin{aligned}
\rho_{{\rm BI}}=\sum_{i}P_{i}\sigma P_{i}=\Delta[\sigma],~\sigma\in\mathcal{S},
\end{aligned}
\end{equation}
where $\Delta$ denotes the block-dephasing operation. The set of block-incoherent states is denoted as $\mathcal{I}_{{\rm B}}(\mathcal{H})$.

 We refer to the largest class of (free) operations that cannot produce block-coherence as maximally block-incoherent (MBI) operations. A channel $\Lambda_{\mathrm{MBI}}$ on $\mathcal{S}$ is an element of this operation class if and only if it maps any block-incoherent state to a block-incoherent state \cite{4,19,20,21}, namely
\begin{equation}\label{7}
\begin{aligned}
\Lambda_{{\rm MBI}}(\mathcal{I}_{{\rm B}}(\mathcal{H}))\subseteq\mathcal{I}_{{\rm B}}(\mathcal{H}),
\end{aligned}
\end{equation}
or equivalently
\begin{equation}\label{7}
\begin{aligned}
\Lambda_{{\rm MBI}}\circ\Delta=\Delta\circ\Lambda_{{\rm MBI}}\circ\Delta.
\end{aligned}
\end{equation}
A quantum channel $\Lambda$ is often expressed by the Kraus operators. Let $\{K_{n}\}$ be a set of Kraus operators on $\mathcal{H}$, and the operators satisfy the normalization condition $\sum_{n}K_{n}^{\dagger}K_{n}=\mathbb{I}$. Some Kraus operators have the form
\begin{equation}\label{7}
\begin{aligned}
K_{n}=\sum_{i}P_{f(i)}c_{n}P_{i},
\end{aligned}
\end{equation}
where $f$ is the index function, $c_{n}$ is the complex matrix. $K_{n}$ is a block-incoherent Kraus operator, if $f$ is an index permutation.

  A real-valued function $\mathcal{C}(\rho,\mathbf{P})$ is called block-coherence monotone of quantum state $\rho$ with  respect to the projective measurement $\mathbf{P}$, if it satisfies \cite{4,19,20,21}:

(B1) Faithfulness: $\mathcal{C}(\rho,\mathbf{P})\geq0$ with equality if $\rho\in\mathcal{I}_{{\rm B}}(\mathcal{H})$.

(B2) Monotonicity: $\mathcal{C}(\Lambda_{{\rm BI}}(\rho),\mathbf{P})\leq\mathcal{C}(\rho,\mathbf{P})$ for any block-incoherent operation $\Lambda_{{\rm BI}}$.

(B3) Strong monotonicity: $\sum_{n}p_{n}\mathcal{C}(\rho_{n},\mathbf{P})\leq\mathcal{C}(\rho,\mathbf{P})$ for any block-incoherent operation $\Lambda_{{\rm BI}}=\{K_n\}$, where $p_{n}={\rm Tr}(K_{n}\rho K_{n}^{\dagger})$, $\rho_{n}=\frac{K_{n}\rho K_{n}^{\dagger}}{p_{n}}$.

(B4) Convexity: $\mathcal{C}(\sum_{n}p_{n}\rho_{n},\mathbf{P})\leq\sum_{n}p_{n}\mathcal{C}(\rho_{n},\mathbf{P})$ for all states $\rho_{n}$, and the probability  $\{p_{n}\} $ which satisfies $p_{n}\geq0$, $\sum_{n}p_{n}=1$.

Note that the rank of the above projector $P_{i}$ is arbitrary, and when the rank of $P_{i}$ is 1, it is consistent with the standard  resource theory of coherence.

\subsection{POVM coherence theoretical framework}\label{sec:review}

The most general quantum measurement refers to  the positive-operator-valued measures (POVM) \cite{19}. Let a set $\mathbf{E}=\{E_{i}\}_{i=1}^{n}$  of positive-semidefinite operators be a POVM on a $d$-dimensional Hilbert space $\mathcal{H}$, and $\sum_{i}E_{i}=\mathbb{I}$, where $E_{i}$ is called POVM element. Suppose $E_{i}=A_{i}^{\dagger}A_{i}$ for any $i$, where $\{A_{i}\}$ is a set of measurement operators for $\mathbf{E}$, and $A_{i}$ can be written as $A_{i}=U_{i}\sqrt{E_{i}}$ with any unitary operator $U_{i}$. The $i$th post-measurement state for a given $A_{i}$ is $\rho_{i}=\frac{A_{i}\rho A_{i}^{\dagger}}{{\rm Tr}[A_{i}\rho A_{i}^{\dagger}]}$ \cite{19,20,22}.

The POVM coherence resource theory is established  via the Naimark extension \cite{19,23}. Every POVM $\mathbf{E}=\{E_{i}\}_{i=1}^{n}$ on a $d$-dimensional Hilbert space $\mathcal{H}$, can be extended to a projective measurement $\mathbf{P}=\{P_{i}\}_{i=1}^{n}$ on the Hilbert space $\mathcal{H'}$, if one can embed the $d$-dimensional Hilbert space $\mathcal{H}$ into a larger $d'$-dimensional Hilbert space $\mathcal{H'}$ called the Naimark space, where $d'\geq d$. The general way to embed the original space $\mathcal{H}$ into a larger space $\mathcal{H}'$ is via a direct sum, namely, in  the Naimark space $\mathcal{H'}$, the corresponding state  $\varepsilon(\rho)$ of quantum state $\rho$  in the $d$-dimensional Hilbert space $\mathcal{H}$ is
\begin{equation}\label{7}
\begin{aligned}
\varepsilon(\rho)=\rho\oplus0,
\end{aligned}
\end{equation}
requiring
\begin{equation}\label{6}
\begin{aligned}
{\rm Tr}[E_{i}\rho]={\rm Tr}[P_{i}\varepsilon(\rho)]={\rm Tr}[P_{i}(\rho\oplus0)]
\end{aligned}
\end{equation}
to hold for all states $\rho$ in a set $\mathcal{S}$ of quantum states. Here $\oplus$ denotes the orthogonal direct sum, and $0$ is zero matrix of dimension $d'-d$. Any projective measurement $\mathbf{P}$ that satisfies Eq. (6) is called a Naimark extension of $\mathbf{E}$.

The  embedding into a larger-dimensional Hilbert space can also be realized via the canonical Naimark extension \cite{19,23}: one  attaches a probe, initially in the state $|1\rangle\langle 1|$, via the tensor product $\varepsilon(\rho)=\rho\otimes|1\rangle\langle1|$ \cite{19}. A canonical Naimark extension projective measurement $\mathbf{P}=\{P_{i}\}_{i=1}^{n}$ of the POVM $\mathbf{E}=\{E_{i}\}_{i=1}^{n}$ is described by a unitary matrix $V$ which makes \cite{19,20}
\begin{equation}\label{7}
\begin{aligned}
P_{i}:=V^{\dagger}(\mathbb{I}\otimes|i\rangle\langle i|)V,
\end{aligned}
\end{equation}
and
\begin{equation}\label{7}
\begin{aligned}
{\rm Tr}[E_{i}\rho]={\rm Tr}[P_{i}(\rho\otimes|1\rangle\langle1|)]
\end{aligned}
\end{equation}
to hold for all states $\rho$ in the quantum state set $\mathcal{S}$.

  A state $\rho$ is called a POVM-incoherent state \cite{19,20,22}, if
\begin{equation}\label{7}
\begin{aligned}
E_{i}\rho E_{j}=0,~~{\rm for~all}~i\neq j,
\end{aligned}
\end{equation}
or equivalently
\begin{equation}\label{7}
\begin{aligned}
A_{i}\rho A_{j}^{\dagger}=0,~~{\rm for~all}~i\neq j.
\end{aligned}
\end{equation}
The set of POVM-incoherent states is denoted as $\mathcal{I}_{{\rm PI}}$.

 A channel $\Lambda$ is called a POVM-incoherent operation (${\rm PI}$) with respect to the POVM $\mathbf{E}=\{E_{i}\}_{i=1}^{n}$, if it admits a Kraus decomposition $\Lambda(\rho)=\sum_{l}K_{l}\rho K_{l}^{\dagger}$ such that all operators $K_{l}$ with respect to a canonical Naimark extension  projective measurement  $\mathbf{P}=\{P_{i}\}_{i=1}^{n}$  of the POVM $\mathbf{E}=\{E_{i}\}_{i=1}^{n}$ satisfies
\begin{equation}\label{7}
\begin{aligned}
K_{l}\rho K_{l}^{\dagger}\otimes|1\rangle\langle1|=K_{l}'(\rho\otimes|1\rangle\langle1|)(K_{l}')^{\dagger},
\end{aligned}
\end{equation}
for all $l\in\{1,2,\ldots,n\}$, where $\{K_{l}'\}$ is a set of the block-incoherent operators on the extended Hilbert space $\mathcal{H}'$ \cite{20}.

 The POVM-based coherence measure $\mathcal{C}(\rho,\mathbf{E})$ of a state $\rho$ with respect to a POVM $\mathbf{E}=\{E_{i}\}_{i=1}^{n}$ is defined as the block-coherence measure $\mathcal{C}(\varepsilon(\rho),\mathbf{P})$ of $\varepsilon(\rho)$ with respect to the Naimark extension $\mathbf{P}$ of $\mathbf{E}$ \cite{19,20,22}, namely
\begin{equation}\label{7}
\begin{aligned}
\mathcal{C}(\rho,\mathbf{E}):=\mathcal{C}(\varepsilon(\rho),\mathbf{P}),
\end{aligned}
\end{equation}
where the function $\mathcal{C}$ on the right side denotes any unitary-covariant block-coherence measure.

The POVM-based coherence measure $\mathcal{C}(\rho,\mathbf{E})$ with respect to a general quantum measurement $\mathbf{E}=\{E_{i}\}_{i=1}^{n}$ should satisfy:

(P1) Faithfulness: $\mathcal{C}(\rho,\mathbf{E})\geq0$ with equality if $\rho\in\mathcal{I}_{{\rm PI}}$.

(P2) Monotonicity: $\mathcal{C}(\Lambda_{{\rm PI}}(\rho),\mathbf{E})\leq\mathcal{C}(\rho,\mathbf{E})$ for any POVM-incoherent operation $\Lambda_{{\rm PI}}$.

(P3) Strong monotonicity: $\sum_{l}p_{l}\mathcal{C}(\rho_{l},\mathbf{E})\leq\mathcal{C}(\rho,\mathbf{E})$ for any POVM-incoherent operation $\Lambda_{{\rm PI}}=\{K_l\}$, where $p_{l}={\rm Tr}(K_{l}\rho K_{l}^{\dagger})$, $\rho_{l}=\frac{K_{l}\rho K_{l}^{\dagger}}{p_{l}}$.

(P4) Convexity: $\mathcal{C}(\sum_{i}p_{i}\rho_{i},\mathbf{E})\leq\sum_{i}p_{i}\mathcal{C}(\rho_{i},\mathbf{E})$ for all states $\rho_{i}$, and the probability $\{p_{i}\}$ satisfing $p_{i}\geq0$, $\sum_{i}p_{i}=1$.

\subsection{The max-relative entropy and the coherent rank}\label{sec:review}

In the theoretical framework of BCP, the max-relative entropy between quantum state $\rho\geq0$ and quantum state $\sigma\geq0$ is defined as \cite{24,25}
\begin{equation}\label{7}
\begin{aligned}
D_{{\rm max}}(\rho\|\sigma)=\log_{2}\min\{\lambda|\rho\leq\lambda\sigma\}.
\end{aligned}
\end{equation}
One equivalent definition of $D_{{\rm max}}(\rho\|\sigma)$ \cite{25} is
\begin{equation}\label{7}
\begin{aligned}
D_{{\rm max}}(\rho\|\sigma):=\log_{2}\min\{\lambda|{\rm Tr}[P_{+}^{\lambda}(\rho-\lambda\sigma)]=0\},
\end{aligned}
\end{equation}
where $P_{+}^{\lambda}$ is the  projector of $\rho-\lambda\sigma$ with positive eigenvalues.

 The coherent rank $\mathcal{C}_{r}$ of a pure state $|\varphi\rangle=\sum_{i=1}^{R}\varphi_{i}|i\rangle$ (not necessarily normalized) with $\varphi_{i}\neq0$ is defined as the number of terms with $\varphi_{i}\neq0$ \cite{7,26}, i.e.,
\begin{equation}\label{7}
\begin{aligned}
\mathcal{C}_{r}(\varphi)=R.
\end{aligned}
\end{equation}

\section{The block-coherence measures}

Based on the max-relative entropy, we first define a block-coherence measure, which is a generalization  of the coherence measure in Ref. \cite{17}.
\begin{definition}The block-coherence measure $\mathcal{C}_{{\rm max}}(\rho,\mathbf{P})$ of a quantum state $\rho$ with respect to the projective measurement $\mathbf{P}$ is defined as
\begin{equation}\label{7}
\begin{aligned}
\mathcal{C}_{{\rm max}}(\rho,\mathbf{P})=\min_{\sigma\in\mathcal{I}_{B}(\mathcal{H})}D_{{\rm max}}(\rho\|\sigma).
\end{aligned}
\end{equation}
\end{definition}

Then, we have the following result.
\begin{proposition}\label{1} The block-coherence measure  $\mathcal{C}_{{\rm max}}(\rho,\mathbf{P})$ is a block-coherence monotone under ${\rm MBI}$ operations and it is quasi-convex.
\end{proposition}

\textit{\textbf{Proof.}} First, we show that $\mathcal{C}_{{\rm max}}(\rho,\mathbf{P})\geq0$ with the equality  if and only if $\rho\in\mathcal{I}_{{\rm B}}(\mathcal{H})$.

By the definition, we known \cite{24,25}
\begin{equation}\label{7}
\begin{aligned}
\mathcal{C}_{{\rm max}}(\rho,\mathbf{P})
&=\min_{\sigma\in\mathcal{I}_{B}(\mathcal{H})}D_{{\rm max}}(\rho\|\sigma)\\
&=\min_{\sigma\in\mathcal{I}_{B}(\mathcal{H})}\log_{2}\min\{\lambda|\rho\leq\lambda\sigma\}.
\end{aligned}
\end{equation}
Since $\rho\leq\lambda\sigma$, we have ${\rm Tr}(\lambda\sigma-\rho)\geq0$. So $\lambda\geq1$ holds. Hence
\begin{equation}\label{7}
\begin{aligned}
\mathcal{C}_{{\rm max}}(\rho,\mathbf{P})\geq0.
\end{aligned}
\end{equation}
From Ref. \cite{24,25}, we know that $D_{{\rm max}}(\rho\|\sigma)=0$ if and only if $\rho=\sigma$. Then, when
\begin{equation}\label{7}
\begin{aligned}
\mathcal{C}_{{\rm max}}(\rho,\mathbf{P})=\min_{\sigma\in\mathcal{I}_{B}(\mathcal{H})}D_{{\rm max}}(\rho\|\sigma)=0,
\end{aligned}
\end{equation}
$\rho$ must be a block-incoherent state. This implies that $\mathcal{C}_{{\rm max}}(\rho,\mathbf{P})$ satisfies (B1).

Second, we can prove that for any MBI operation with $\{K_{n}\}$, $\mathcal{C}_{{\rm max}}(\rho,\mathbf{P})$ satisfies (B2). According to Ref. \cite{24}, we know that the max-relative entropy $D_{{\rm max}}(\rho\|\sigma)$ are monotonic under completely positive trace-preserving map (CPTP) $\Lambda$. Hence
\begin{equation}\label{8}
\begin{aligned}
D_{{\rm max}}(\Lambda(\rho)\|\Lambda(\sigma))\leq D_{{\rm max}}(\rho\|\sigma).
\end{aligned}
\end{equation}
As any MBI operation $\Lambda_{{\rm MBI}}$ with $\{K_{n}\}$ is a CPTP map, we have
\begin{equation}\label{8}
\begin{aligned}
\min_{\sigma\in\mathcal{I}_{B}(\mathcal{H})}D_{{\rm max}}(\Lambda_{{\rm MBI}}(\rho)\|\Lambda_{{\rm MBI}}(\sigma))\leq\min_{\sigma\in\mathcal{I}_{B}(\mathcal{H})}D_{{\rm max}}(\rho\|\sigma).
\end{aligned}
\end{equation}
Therefore,
\begin{equation}\label{8}
\begin{aligned}
\mathcal{C}_{{\rm max}}(\sum_{n}K_{n}\rho K_{n}^{\dag},\mathbf{P})\leq\mathcal{C}_{{\rm max}}(\rho,\mathbf{P}).
\end{aligned}
\end{equation}
It means that $\mathcal{C}_{{\rm max}}(\rho,\mathbf{P})$ satisfies (B2).

Next we will show that $\mathcal{C}_{{\rm max}}(\rho,\mathbf{P})$ is quasi-convex, i.e.,
\begin{equation}\label{8}
\begin{aligned}
\mathcal{C}_{{\rm max}}(\sum_{n}p_{n}\rho_{n},\mathbf{P})\leq\max_{n}\mathcal{C}_{{\rm max}}(\rho_{n},\mathbf{P}),
\end{aligned}
\end{equation}
where $p_{n}={\rm Tr}(K_{n}\rho K_{n}^{\dag}),~\rho_{n}=\frac{K_{n}\rho K_{n}^{\dag}}{p_{n}}$.

In order to prove the above conclusion we need the following result: For self-adjoint operators $A, B$ and any positive operator $0\leq P\leq I$, we have \cite{25,27,28}
\begin{equation}\label{9}
\begin{aligned}
{\rm Tr}[P(A-B)]\leq{\rm Tr}[\{A\geq B\}(A-B)],\\
{\rm Tr}[P(A-B)]\geq{\rm Tr}[\{A\leq B\}(A-B)].
\end{aligned}
\end{equation}

Now let's prove that $\mathcal{C}_{{\rm max}}(\rho,\mathbf{P})$ is quasi-convex. For any mixture of states, $\rho=\sum p_{n}\rho_{n}$, we can construct a block-incoherent state $\sigma=\sum p_{n}\sigma_{n}$, where every $\sigma_{n}$ is a block-incoherent state. Another equivalent definition of the max-relative entropy $D_{{\rm max}}(\rho\|\sigma)$ \cite{24} is
\begin{equation}\label{8}
\begin{aligned}
D_{{\rm max}}(\rho\|\sigma):=\log_{2}\min\{\lambda|{\rm Tr}[P_{+}^{\lambda}(\rho-\lambda\sigma)]=0\},
\end{aligned}
\end{equation}
where $P_{+}^{\lambda}$ is the  projector of $\rho-\lambda\sigma$ with positive eigenvalues. By (24), we have \cite{25}
\begin{equation}\label{9}
\begin{aligned}
0\leq{\rm Tr}[P_{+}^{\lambda}(\rho-\lambda\sigma)]
&=\sum_{n}p_{n}{\rm Tr}[P_{+}^{\lambda}(\rho_{n}-\lambda\sigma_{n})]\\
&\leq\sum_{n}p_{n}{\rm Tr}[P_{+}^{\lambda,n}(\rho_{n}-\lambda\sigma_{n})],
\end{aligned}
\end{equation}
 where $P_{+}^{\lambda,n}$ is the  projector of $\rho_{n}-\lambda\sigma_{n}$ with positive eigenvalues. Set $\lambda=\max \lambda_{n}$, where for each $n$, $\lambda_{n}$ is defined by $\log_{2}\lambda_{n}=\mathcal{C}_{{\rm max}}(\rho_{n},\mathbf{P})$.

For this choice of $\lambda$, there is ${\rm Tr}[P_{+}^{\lambda}(\rho-\lambda\sigma)]=0$, and hence $\log_{2}\lambda\geq\mathcal{C}_{{\rm max}}(\rho,\mathbf{P})$, i.e.,
\begin{equation}\label{10}
\begin{aligned}
\mathcal{C}_{{\rm max}}(\sum_{n}p_{n}\rho_{n},\mathbf{P})\leq\max_{n}\mathcal{C}_{{\rm max}}(\rho_{n},\mathbf{P}).
\end{aligned}
\end{equation}
So $\mathcal{C}_{{\rm max}}(\rho,\mathbf{P})$ is the quasi-convex.~$\hfill\blacksquare$
\\
\\
\begin{definition} A maximally block-coherent state is defined by
\begin{equation}\label{6}
\begin{aligned}
|\psi_{N}\rangle=\frac{1}{\sqrt{N}}\sum_{k=1}^{N}\frac{P_{k}|\psi_{d}\rangle}{\sqrt{\langle\psi_{d}|P_{k}|\psi_{d}\rangle}},
\end{aligned}
\end{equation}
where $|\psi_{d}\rangle=\frac{1}{\sqrt{d}}\sum_{i=1}^{d}|i\rangle$ is maximally coherent state in the $d$-dimensional Hilbert space, and the rank of projective measurement $P_{k}$ is arbitrary and the number of $P_{k}$ in the projective measurement ${\bf P}=\{P_{k}\}$ is $N$~($N\leq d$).
\end{definition}

Obviously, for a maximally block-coherent state $|\psi_{N}\rangle$, we have
\begin{equation}\label{10}
\begin{aligned}
\mathcal{C}_{{\rm max}}(\psi_{N},\mathbf{P})=\log_{2}N,
\end{aligned}
\end{equation}
namely the value of $\mathcal{C}_{{\rm max}}(\psi_{N},\mathbf{P})$ depends on the number $N$ of projectors in the space.

 One-shot scenario is the most general conversion case, where the conversion is from an initial state to a  final state. One-shot  block coherence dilution process is to convert the maximally block-coherent state $|\psi_{N}\rangle$ into the desired state $\rho$  via the maximally block-incoherent operation \cite{13,17,18}.

First, we define a block-coherence measure, the one-shot block coherence cost of quantifying block coherence dilution.

\begin{definition} Let ${\rm MBI}$ denote the set of the  maximally block-incoherent operations. For a given state $\rho$ and $\epsilon\geq 0$, the one-shot block coherence cost under the maximally block-incoherent operations is defined as
\begin{equation}\label{6}
\begin{aligned}
\mathcal{C}^{\epsilon}_{{\rm MBI}}(\rho,\mathbf{P})=\min_{\Lambda\in {\rm MBI}}\{\log_{2}N|F[\Lambda(|\psi_{N}\rangle),\rho]\geq1-\epsilon\},
\end{aligned}
\end{equation}
while $F(\rho,\sigma)=({\rm Tr}[\sqrt{\sqrt{\rho}\sigma\sqrt{\rho}}])^{2}$ is the fidelity between two quantum states $\rho$ and $\sigma$, where $|\psi_{N}\rangle$ is the maximally block-coherent state.
\end{definition}

Since the one-shot scenario allows errors to exist, in the presence of the error $\epsilon$, we use
\begin{equation}\label{5}
\begin{aligned}
C^{\epsilon}(\rho):=\min_{\rho':F(\rho,\rho')\geq1-\epsilon}C(\rho')
\end{aligned}
\end{equation}
to characterize the coherence measure of state $\rho$ \cite{17}. That is in order to define the coherence cost with a certain error $\epsilon$, one can use a smoothing to the measure $C(\rho)$ by  minimizing over states $\rho'$ satisfying $F(\rho,\rho')\geq1-\epsilon$ to smooth the  measure $\mathcal{C}(\rho)$.

Next we discuss the relationship between the coherence measure $\mathcal{C}^{\epsilon}_{{\rm max}}(\rho,\mathbf{P})$ and the one-shot block coherence cost $\mathcal{C}^{\epsilon}_{{\rm MBI}}(\rho,\mathbf{P})$. We have

\begin{theorem}\label{1} For $\epsilon>0$, the coherence measures satisfy
\begin{equation}\label{25}
\begin{aligned}
\mathcal{C}^{\epsilon}_{{\rm max}}(\rho,\mathbf{P})\leq\mathcal{C}^{\epsilon}_{{\rm MBI}}(\rho,\mathbf{P})\leq\mathcal{C}^{\epsilon}_{{\rm max}}(\rho,\mathbf{P})+1.
\end{aligned}
\end{equation}
\end{theorem}

\textit{\textbf{Proof.}} We first prove the left side of Eq.~(32)~. For quantum state $\rho$ and the projective measurement $\mathbf{P}=\{P_i\}$, let $\log_{2}N'=\mathcal{C}^{\epsilon}_{{\rm MBI}}(\rho,\mathbf{P})$, where the rank of $P_{i}$ is arbitrary. The definition of $\mathcal{C}^{\epsilon}_{{\rm MBI}}(\rho,\mathbf{P})$ means that there is an operation $\Lambda_{{\rm MBI}}$ which makes $F[\Lambda_{{\rm MBI}}(|\psi_{N'}\rangle),\rho]\geq1-\epsilon$ true, then
\begin{equation}\label{26}
\begin{aligned}
\mathcal{C}^{\epsilon}_{{\rm max}}(\rho,\mathbf{P})
&\leq\mathcal{C}_{{\rm max}}(\Lambda_{{\rm MBI}}(\psi_{N'}),\mathbf{P})\\
&=\min_{\delta\in\mathcal{I}_{B}(\mathcal{H})}D_{{\rm max}}(\Lambda_{{\rm MBI}}(\psi_{N'})\|\delta)\\
&\leq D_{{\rm max}}(\Lambda_{{\rm MBI}}(\psi_{N'})\|\Lambda_{{\rm MBI}}(\sigma))\\
&\leq D_{{\rm max}}(\psi_{N'}\|\sigma)\\
&=\log_{2}\min\{\lambda|\psi_{N'}\leq\lambda\sigma\}.
\end{aligned}
\end{equation}
Here $\sigma$ is a block-incoherent state and $\psi_{N'}=|\psi_{N'}\rangle\langle \psi_{N'}|$. We calculate the critical value of $\lambda$ in the case of $\psi_{N'}=\lambda\sigma$. It is easy to get $\lambda=N'$. Hence
\begin{equation}\label{27}
\begin{aligned}
\mathcal{C}^{\epsilon}_{{\rm max}}(\rho,\mathbf{P})\leq D_{{\rm max}}(\psi_{N'}\|\sigma)=\log_{2}N'=\mathcal{C}^{\epsilon}_{{\rm MBI}}(\rho,\mathbf{P}).
\end{aligned}
\end{equation}
So
\begin{equation}\label{28}
\begin{aligned}
\mathcal{C}^{\epsilon}_{{\rm max}}(\rho,\mathbf{P})\leq\mathcal{C}^{\epsilon}_{{\rm MBI}}(\rho,\mathbf{P}).
\end{aligned}
\end{equation}

Next we prove the right side of Eq.~(32)~. Assume that the state $\rho'$ reaches minimum, so
\begin{equation}\label{29}
\begin{aligned}
\mathcal{C}^{\epsilon}_{{\rm max}}(\rho,\mathbf{P})
&=\mathcal{C}_{{\rm max}}(\rho',\mathbf{P})\\
&=D_{{\rm max}}(\rho'\|\delta)\\
&=\log_{2}\lambda.
\end{aligned}
\end{equation}
Set $N''=\lceil\lambda\rceil$, then $\rho'\leq N''\delta$. Consider the following mapping
\begin{equation}\label{30}
\begin{aligned}
\Lambda(\omega)
&=\frac{1}{N''-1}(N''{\rm Tr}[\psi_{N''}\circ\omega]-1)\rho'\\
&~~~+\frac{N''}{N''-1}(1-{\rm Tr}[\psi_{N''}\circ\omega])\delta,
\end{aligned}
\end{equation}
where $\psi_{N''}\circ\omega=|\psi_{N''}\rangle\langle\psi_{N''}|\omega$, ${\rm Tr}[\psi_{N''}\circ\omega]=\langle\psi_{N''}|\omega|\psi_{N''}\rangle$. For all $\delta=\sum_{i=1}P_{i}\delta P_{i}\in\mathcal{I}_{B}(\mathcal{H})$, we have ${\rm Tr}[\psi_{N''}\circ\delta]=\frac{1}{N''}$ and  $\Lambda(\delta)=\delta\in\mathcal{I}_{B}(\mathcal{H})$. So $\Lambda\in {\rm MBI}$. On the other hand, it is easy to obtain $\Lambda(\psi_{N''})=\rho'$. One can also write the mapping as
\begin{equation}\label{31}
\begin{aligned}
&~~~~\Lambda(\omega)\\
&=\frac{N''}{N''-1}{\rm Tr}[\psi_{N''}\circ\omega]\rho'-\frac{1}{N''-1}\rho'\\
&~~~+\frac{N''}{N''-1}(1-{\rm Tr}[\psi_{N''}\circ\omega])\delta\\
&=\frac{1}{N''-1}{\rm Tr}[\psi_{N''}\circ\omega]\rho'-\frac{1}{N''-1}\rho'+{\rm Tr}[\psi_{N''}\circ\omega]\rho'\\
&~~~+\frac{N''}{N''-1}(1-{\rm Tr}[\psi_{N''}\circ\omega])\delta\nonumber\\
&=\frac{N''}{N''-1}({\rm Tr}[\psi_{N''}\circ\omega]-1)\frac{1}{N''}\rho'\\
\end{aligned}
\end{equation}
\begin{equation}
\begin{aligned}
&~~~+\frac{N''}{N''-1}(1-{\rm Tr}[\psi_{N''}\circ\omega])\delta+{\rm Tr}[\psi_{N''}\circ\omega]\rho'\\
&=\frac{N''}{N''-1}(1-{\rm Tr}[\psi_{N''}\circ\omega])(\delta-\frac{1}{N''}\rho')+{\rm Tr}[\psi_{N''}\circ\omega]\rho'.
\end{aligned}
\end{equation}

Since $\delta\geq\frac{1}{N''}\rho'$, $\Lambda$ is an entirely positive maximally block-incoherent operation, which maps $|\psi_{N''}\rangle$ to $\rho'$ so we have
\begin{equation}\label{32}
\begin{aligned}
\mathcal{C}^{\epsilon}_{{\rm MBI}}(\rho,\mathbf{P})=\log_{2}N''\leq\log_{2}(1+\lambda)\\
\leq\log_{2}\lambda+1=\mathcal{C}^{\epsilon}_{{\rm max}}(\rho,\mathbf{P})+1.
\end{aligned}
\end{equation}
Therefore, the Eq.~(32)~holds. ~$\hfill\blacksquare$
\\

 Now, we discuss the deterministic block  coherence dilution between the maximally block-coherent state and the pure  block-coherent  state under block-incoherent ({\rm BI}) operation, by the method of construction of block-incoherent operator \cite{18,29,30,31,32}.

Assume that  $|\phi\rangle=\sum_{i}\phi_{i}|i\rangle$ is a  block-coherent pure state in the quantum state set $\mathcal{S}$, where $\{\phi_{i}
\}_{i=1, 2, \cdots}$  are non-negative real numbers, and all complex phases have been eliminated by block-incoherent operations. The block coherence dilution process is $|\psi_{N}\rangle\stackrel{{\rm BI}}{\longrightarrow}|\phi\rangle$, where $|\psi_{N}\rangle=\frac{1}{\sqrt{N}}\sum_{k=1}^{N}\frac{P_{k}|\psi_{d}\rangle}{\sqrt{\langle\psi_{d}|P_{k}|\psi_{d}\rangle}}$ is a maximally block-coherent states, and the BI stands for block-incoherent operations. Then, the deterministic block coherence dilution can be described as
\begin{equation}\label{32}
\begin{aligned}
\Lambda_{\rm BI}(|\psi_{N}\rangle\langle\psi_{N}|)=\sum_{n}K_{n}|\psi_{N}\rangle\langle\psi_{N}|K_{n}^{\dagger}=|\phi\rangle\langle\phi|.
\end{aligned}
\end{equation}
Here $\Lambda_{\rm BI}$ is a block-incoherent operation composed by the Kraus operators  $\{K_{n}\}$.
We choose the block-incoherent Kraus operators \cite{22} being
\begin{equation}\label{32}
\begin{aligned}
K_{n}=\sum_{i}P_{\pi(i)}c_{n}^{i}P_{i},
\end{aligned}
\end{equation}
where $P_{\pi(i)}$ is the $\pi(i)$-th projective measurement, $\pi$ is the element of the set of permutation, and $c_{n}^{i}$ is the complex number. The Kraus operator satisfies the condition $\sum_{n}(K_{n})^{\dagger}(K_{n})=\mathbb{I}$.
So we can rewrite the Eq.~(40) as

\begin{equation}\label{32}
\begin{aligned}
&~~~~\Lambda_{\rm BI}(|\psi_{N}\rangle\langle\psi_{N}|)\\
&=\sum_{i,j,n}P_{\pi(i)}c_{n}^{i}P_{i}|\psi_{N}\rangle\langle\psi_{N}|P_{j}c_{n}^{j*}P_{\pi(j)}\\
&=\sum_{i,j,n}P_{\pi(i)}c_{n}^{i}c_{n}^{j*}\sum_{k,l=1}^{N}\frac{P_{i}P_{k}|i\rangle\langle j|P_{l}P_{j}}{Nd\sqrt{\langle\psi_{d}|P_{k}|\psi_{d}\rangle}\sqrt{\langle\psi_{d}|P_{l}|\psi_{d}\rangle}}P_{\pi(j)}\\
&=\sum_{i,j}\phi_{i}\phi_{j}^{*}|i\rangle\langle j|.
\end{aligned}
\end{equation}

For the sake of simplicity  here we only discuss the case of $N=d$. In this case,  deterministic coherence dilution means that the coefficients between the initial state $|\psi_{d}\rangle$ (maximally block-coherent state) and the final state $|\phi\rangle$ satisfy the majorization relation \cite{29}, i.e.,
\begin{equation}\label{32}
\begin{aligned}
&~~~\mu(\psi_{d})=(||\psi_{d}\rangle|^{2},||\psi_{d}\rangle|^{2},\ldots,||\psi_{d}\rangle|^{2})^{T}\\
&\prec\mu(\phi)=(\phi_{1}^{2},\phi_{2}^{2},\ldots,\phi_{d}^{2})^{T},
\end{aligned}
\end{equation}
where $||\psi_{d}\rangle|^{2}=\frac{1}{d}$.

According to the protocol \cite{30} for the deterministic transformations of the coherent states  for which the majorization relation can be satisfied,  for  the case $\frac {1}{\sqrt{d}}\geq \phi_d$, $\frac {1}{\sqrt{d}}\leq  \phi_i, i=1,2,\ldots, d-1$,   the set of $d$ permutations in that case turns out to be
\begin{equation}\label{32}
\begin{aligned}
&~~~\{U^i|i=1,2,\ldots, d\}\\
&=\{I_{d},|1\rangle\leftrightarrow|d\rangle,|2\rangle\leftrightarrow|d\rangle,\ldots,|d-1\rangle\leftrightarrow|d\rangle\};
\end{aligned}
\end{equation}
the probabilities in that case turn out to be \cite{30}
\begin{equation}\label{32}
\begin{aligned}
p^{1}=1-\sum_{i=2}^{d}p^{i},~~p^{i}=\frac{\phi_{i-1}^{2}-\psi_{i-1}^{2}}{\phi_{i-1}^{2}-\phi_{d}^{2}};
\end{aligned}
\end{equation}
the set of Kraus operators of the incoherent operation \cite{30} is
\begin{equation}
\{K^i=U^i\sqrt {p^i}\sum_{j=1}^dc_{ij}{\sqrt d}|j\rangle\langle j|, i=1,2,\ldots, d\},
\end{equation}
where $c_{ij}$ is the $(ij)$th element of the $d\times d$ matrix $c$  satisfying
\begin{equation}
U^i(c_{i1}, c_{i2}, \cdots, c_{id})^T=(\phi_1,\phi_2,\cdots, \phi_d)^T.
\end{equation}

For example, when $d=4, N=4$, we have  $P_{1}=|1\rangle\langle1|$, $P_{2}=|2\rangle\langle2|$, $P_{3}=|3\rangle\langle3|$ and $P_{4}=|4\rangle\langle 4|$. The maximally block-coherent state $|\psi_{4}\rangle=\frac{\sum_{k=1}^{4}\sum_{i=1}^{4}P_{k}|i\rangle}{4\sqrt{\langle\psi_{4}|P_{k}|\psi_{4}\rangle}}=\frac {1}{2}\sum_{i=1}^4|i\rangle$. We choose the dilution state $|\phi\rangle=\sum_i \phi_i|i\rangle= \sqrt {0.4}|1\rangle+\sqrt {0.3}|2\rangle+\sqrt {0.28}|3\rangle+\sqrt {0.02}|4\rangle$. So  $\mu(\phi)=(0.4,0.3,0.28,0.02)$ and $\mu(\psi_{4})=(0.25,0.25,0.25,0.25)$. It is easy to see
\begin{equation}\label{32}
\begin{aligned}
&~~~~\mu(\psi_{4})=(0.25,0.25,0.25,0.25)\\
&\prec\mu(\phi)=(0.4,0.3,0.28,0.02).
\end{aligned}
\end{equation}
Obviously $|\phi\rangle$ and $|\psi_4\rangle$ satisfy $\frac {1}{\sqrt{d}}\geq \phi_d$, $\frac {1}{\sqrt{d}}\leq  \phi_i, i=1,2,\ldots, d-1$, when $d=4$. The set of permutations for this case should be
\begin{equation}\label{32}
\begin{aligned}
\{I_{4},|1\rangle\leftrightarrow|4\rangle,|2\rangle\leftrightarrow|4\rangle,|3\rangle\leftrightarrow|4\rangle\}.
\end{aligned}
\end{equation}
The matrix $c$ corresponding to this  set of permutations is
\begin{equation}       
c=\begin{pmatrix}  
    \phi_{1} & \phi_{2} & \phi_{3} & \phi_{4}\\  
    \phi_{4} & \phi_{2} & \phi_{3} & \phi_{1}\\
    \phi_{1} & \phi_{4} & \phi_{3} & \phi_{2}\\
    \phi_{1} & \phi_{2} & \phi_{4} & \phi_{3}\\
  \end{pmatrix}=                
  \begin{pmatrix}   
    \sqrt{\frac{2}{5}} & \sqrt{\frac{3}{10}} & \frac{\sqrt{7}}{5} & \frac{1}{5\sqrt{2}}\\  
    \frac{1}{5\sqrt{2}} & \sqrt{\frac{3}{10}} & \frac{\sqrt{7}}{5} & \sqrt{\frac{2}{5}}\\
    \sqrt{\frac{2}{5}} & \frac{1}{5\sqrt{2}} & \frac{\sqrt{7}}{5} & \sqrt{\frac{3}{10}}\\
    \sqrt{\frac{2}{5}} & \sqrt{\frac{3}{10}} & \frac{1}{5\sqrt{2}} & \frac{\sqrt{7}}{5}\\
  \end{pmatrix}.
\end{equation}
Thus, the set of probabilities is found to be
\begin{equation}\label{32}
\begin{aligned}
p^{2}=\frac{\phi_{1}^{2}-\psi_{4}^{2}}{\phi_{1}^{2}-\phi_{4}^{2}}=\frac{15}{38},\\
p^{3}=\frac{\phi_{2}^{2}-\psi_{4}^{2}}{\phi_{2}^{2}-\phi_{4}^{2}}=\frac{5}{28},\\
p^{4}=\frac{\phi_{3}^{2}-\psi_{4}^{2}}{\phi_{3}^{2}-\phi_{4}^{2}}=\frac{3}{26},\\
p^{1}=1-p^{2}-p^{3}-p^{4}=\frac{2153}{6916}.
\end{aligned}
\end{equation}

Then the Kraus operators are
\begin{equation}\label{32}
\begin{aligned}
K^{1}
&=U^1\sqrt{p^{1}}(2c_{11}|1\rangle\langle1|+2c_{12}|2\rangle\langle2|+2c_{13}|3\rangle\langle3|\\
&~~~~+2c_{14}|4\rangle\langle4|),\\
K^{2}
&=U^2\sqrt{p^{2}}(2c_{21}|1\rangle\langle1|+2c_{22}|2\rangle\langle2|+2c_{23}|3\rangle\langle3|\\
&~~~~+2c_{24}|4\rangle\langle4|),\\
K^{3}
&=U^3\sqrt{p^{3}}(2c_{31}|1\rangle\langle1|+2c_{32}|2\rangle\langle2|+2c_{33}|3\rangle\langle3|\\
&~~~~+2c_{34}|4\rangle\langle4|),\\
K^{4}
&=U^4\sqrt{p^{4}}(2c_{41}|1\rangle\langle1|+2c_{42}|2\rangle\langle2|+2c_{43}|3\rangle\langle3|\\
&~~~~+2c_{44}|4\rangle\langle4|),
\end{aligned}
\end{equation}
where $U^1=I_4$ is the identity transformation, $U^2=|1\rangle\leftrightarrow|4\rangle|,  U^3=|2\rangle\leftrightarrow|4\rangle,  U^4=|3\rangle\leftrightarrow|4\rangle$.
The Kraus operators can be expressed in the following form
\begin{equation}       
K^{1}=U^1\begin{pmatrix}  
    \sqrt{\frac{4306}{8645}} & 0 & 0 & 0\\  
    0 & \sqrt{\frac{6459}{17290}} & 0 & 0\\
    0 & 0 & \frac{\sqrt{2153}}{5\sqrt{247}} & 0\\
    0 & 0 & 0 & \frac{\sqrt{2153}}{5\sqrt{3458}}\\
  \end{pmatrix},
\end{equation}
\begin{equation}       
K^{2}=U^2\begin{pmatrix}   
    \sqrt{\frac{3}{95}} & 0 & 0 & 0\\  
    0 & \frac{3}{\sqrt{19}} & 0 & 0\\
    0 & 0 & \sqrt{\frac{42}{95}} & 0\\
    0 & 0 & 0 & \frac{2\sqrt{3}}{\sqrt{19}}\\
  \end{pmatrix},
\end{equation}
\begin{equation}       
K^{3}=U^3\begin{pmatrix}   
    \sqrt{\frac{2}{7}} & 0 & 0 & 0\\  
    0 & \frac{1}{\sqrt{70}} & 0 & 0\\
    0 & 0 & \frac{1}{\sqrt{5}} & 0\\
    0 & 0 & 0 & \sqrt{\frac{3}{14}}\\
  \end{pmatrix},
\end{equation}
\begin{equation}       
K^{4}=U^4\begin{pmatrix}  
    \frac{2\sqrt{3}}{\sqrt{65}} & 0 & 0 & 0\\  
    0 & \frac{3}{\sqrt{65}} & 0 & 0\\
    0 & 0 & \frac{\sqrt{3}}{5\sqrt{13}} & 0\\
    0 & 0 & 0 & \frac{\sqrt{42}}{5\sqrt{13}}\\
  \end{pmatrix},
\end{equation}
and they satisfy $\sum_{i=1}^{4}(K^{i})^{\dagger}K^{i}=\mathbb{I}_{4}$, which  is an identity operation.

Next we define another coherence measure based on coherent rank.

\begin{definition} A block-coherence measure based on coherent rank   is defined as
\begin{equation}\label{7}
\begin{aligned}
\mathcal{C}_{{\rm 0}}(\rho,\mathbf{P})=\min_{\{p_{i},|\psi_{i}\rangle\}}\max_{i}\log_{2}M(|\psi_{i}\rangle),
\end{aligned}
\end{equation}
where $\mathbf{P}$ is the projective measurement, the minimum is taken over all possible pure state decompositions $\rho=\sum_i p_i|\psi_i\rangle\langle \psi_i|$ with $p_i\geq 0$ and $\sum_i p_i=1$, $M(|\psi_{i}\rangle)$ is the number of $P_j$ satisfying $\langle\psi_{i}|P_{j}|\psi_{i}\rangle\neq0$.
\end{definition}

We have the following result.

\begin{proposition}\label{2} The quantity $\mathcal{C}_{{\rm 0}}(\rho,\mathbf{P})$
is a coherence monotone under the block-incoherent operation ${\rm BI}$.
\end{proposition}

\textit{\textbf{Proof.}} Apparently  $\mathcal{C}_{{\rm 0}}(\rho,\mathbf{P})\geq0$. Next we prove that the quantity  $\mathcal{C}_{{\rm 0}}(\rho,\mathbf{P})$ satisfies $\mathcal{C}_{{\rm 0}}(\rho,\mathbf{P})=0\Leftrightarrow\rho\in\mathcal{I}_{{\rm B}}(\mathcal{H})$.

Suppose $\mathcal{C}_{{\rm 0}}(\rho,\mathbf{P})=0$ and the corresponding ensemble of $\rho$ is $\{p_{i},|\psi_{i}\rangle\}$, we can deduce for all $i$, $|\psi_{i}\rangle\langle\psi_{i}|=P_{j}|\psi_{i}\rangle\langle\psi_{i}|P_{j}$, which means that $\rho\in\mathcal{I}_{{\rm B}}(\mathcal{H})$. Conversely, suppose $\rho=\sum_{i}P_{i}\delta P_{i}=\sum_{i,j=1}^{N}\delta_{j}P_{i}|\psi_{j}\rangle\langle\psi_{j}|P_{i}$, we can choose $\{\delta_{j},P_{i}|\psi_{j}\rangle\}$ as a decomposition, which leads to $\mathcal{C}_{{\rm 0}}(\rho,\mathbf{P})=0$. Hence $\mathcal{C}_{{\rm 0}}(\rho,\mathbf{P})$ satisfies (B1).

Then, we prove that for any block-incoherent operation with $\{K_{n}\}$, there is
\begin{equation}\label{7}
\begin{aligned}
\mathcal{C}_{{\rm 0}}(\sum_{n}K_{n}\rho K_{n}^{\dagger},\mathbf{P})\leq\mathcal{C}_{{\rm 0}}(\rho,\mathbf{P}).
\end{aligned}
\end{equation}

Before we prove above conclusion, let's introduce the following lemma proved in Ref. \cite{7}.

\begin{lemma}\label{2}  If $|\psi\rangle=\frac{K_{i}|\phi\rangle}{\sqrt{{\rm Tr}[K_{i}|\phi\rangle\langle\phi|K_{i}^{\dag}]}}$, where $\{K_{i}\}$ is a set of incoherent-preserving Kraus operators, then $\mathcal{C}_{{\rm 0}}(|\psi\rangle\langle\psi|)\leqslant\mathcal{C}_{{\rm 0}}(|\phi\rangle\langle\phi|)$.
\end{lemma}

It is easy to see that the Lemma 1 also holds when $\{K_{i}\}$ is a block-incoherent operation.

Let $\{p_{i},|\psi_{i}\rangle\}$ be the decomposition such that
\begin{equation}\label{7}
\begin{aligned}
\mathcal{C}_{{\rm 0}}(\rho,\mathbf{P})=\max_{i}\log_{2}M(|\psi_{i}\rangle).
\end{aligned}
\end{equation}
Let $\Lambda_{{\rm BI}}$ be any block-incoherent operation with $\Lambda_{{\rm BI}}(\rho)=\sum_{n}K_{n}\rho K_{n}^{\dagger}$. For a given state $\rho=\sum_{i} p_{i}|\psi_{i}\rangle\langle\psi_{i}|$, then the post-measurement state of $n$th outcome is
\begin{equation}\label{2}
\begin{aligned}
|\psi_{i}^{n}\rangle=\frac{K_{n}|\psi_{i}\rangle}{\sqrt{{\rm Tr}[K_{n}|\psi_{i}\rangle\langle\psi_{i}|K_{n}^{\dag}]}}.
\end{aligned}
\end{equation}
Then we can get an ensemble $\{p(i|n),|\psi_{i}^{n}\rangle\}$, where the probability is
\begin{equation}\label{2}
\begin{aligned}
\textit{p}(i|n)=p_{i}\frac{{\rm Tr}[K_{n}|\psi_{i}\rangle\langle\psi_{i}|K_{n}^{\dag}]}{{\rm Tr}[K_{n}\rho K_{n}^{\dag}]}.
\end{aligned}
\end{equation}
Then the corresponding density operator $\rho_{n}$ of $n$th outcome is
\begin{equation}\label{2}
\begin{aligned}
\rho_{n}=\frac{\sum_{i}p_{i}K_{n}|\psi_{i}\rangle\langle\psi_{i}|K_{n}^{\dag}}{{\rm Tr}[K_{n}\rho K_{n}^{\dag}]}.
\end{aligned}
\end{equation}
According to the Lemma 1, we can know that for the minimum ensemble decomposition, there is $\mathcal{C}_{{\rm 0}}(\rho_{n},\mathbf{P})\leq\mathcal{C}_{{\rm 0}}(\rho,\mathbf{P})$, and then
\begin{equation}\label{2}
\begin{aligned}
\mathcal{C}_{{\rm 0}}(\sum_{n}K_{n}\rho K_{n}^{\dagger},\mathbf{P})\leq\mathcal{C}_{{\rm 0}}(\rho,\mathbf{P}).\nonumber
\end{aligned}
\end{equation}
This implies $\mathcal{C}_{{\rm 0}}(\rho,\mathbf{P})$ satisfies (B2).~$\hfill\blacksquare$
\\

Now we discuss the relationship between $\mathcal{C}^{\epsilon}_{{\rm MBI}}(\rho,\mathbf{P})$ and $\mathcal{C}^{\epsilon}_{{\rm 0}}(\rho,\mathbf{P})$.    We have the following conclusion.

\begin{theorem}\label{2} For $\epsilon>0$, the value of the one-shot block coherence cost under ${\rm MBI}$ is equal to $\mathcal{C}^{\epsilon}_{{\rm 0}}(\rho,\mathbf{P})$, namely
\begin{equation}\label{25}
\begin{aligned}
\mathcal{C}^{\epsilon}_{{\rm MBI}}(\rho,\mathbf{P})=\mathcal{C}^{\epsilon}_{{\rm 0}}(\rho,\mathbf{P}).
\end{aligned}
\end{equation}
\end{theorem}

\textit{\textbf{Proof.}} First we study the lower bound on $\mathcal{C}^{\epsilon}_{{\rm MBI}}(\rho,\mathbf{P})$. Let $\log_{2}N=\mathcal{C}^{\epsilon}_{{\rm MBI}}(\rho,\mathbf{P})$, then there exists an operation $\Lambda_{{\rm MBI}}$ such that $F[\Lambda_{{\rm MBI}}(\psi_{N}),\rho]\geq1-\epsilon$. Then we have
\begin{equation}\label{25}
\begin{aligned}
\mathcal{C}^{\epsilon}_{{\rm 0}}(\rho,\mathbf{P})
&\leq\mathcal{C}_{{\rm 0}}(\Lambda_{{\rm MBI}}(\psi_{N}),\mathbf{P})\\
&\leq\mathcal{C}_{{\rm 0}}(\psi_{N},\mathbf{P})=\log_{2}N=\mathcal{C}^{\epsilon}_{{\rm MBI}}(\rho,\mathbf{P}).
\end{aligned}
\end{equation}

For the upper bound on $\mathcal{C}^{\epsilon}_{{\rm MBI}}(\rho,\mathbf{P})$, we select the state $\rho'$ reaching minimum such that $\mathcal{C}^{\epsilon}_{{\rm 0}}(\rho,\mathbf{P})=\mathcal{C}_{{\rm 0}}(\rho',\mathbf{P})$. Let $\mathcal{C}_{{\rm 0}}(\rho',\mathbf{P})=\log_{2}N'$, as the {\rm MBI} operation is constructed in the deterministic coherence dilution process discussed above, there is a $\Lambda_{{\rm MBI}}$ such that $F[\Lambda_{{\rm MBI}}(\psi_{N'}),\rho]=F[\rho',\rho]\geq1-\epsilon$, thus
\begin{equation}\label{25}
\begin{aligned}
\mathcal{C}^{\epsilon}_{{\rm MBI}}(\rho,\mathbf{P})
\leq\mathcal{C}_{{\rm MBI}}(\rho',\mathbf{P})=\log_{2}N'=\mathcal{C}_{{\rm 0}}(\rho',\mathbf{P})=\mathcal{C}^{\epsilon}_{{\rm 0}}(\rho,\mathbf{P}).
\end{aligned}
\end{equation}
Therefore, we obtain
\begin{equation}\label{25}
\begin{aligned}
\mathcal{C}^{\epsilon}_{{\rm MBI}}(\rho,\mathbf{P})=\mathcal{C}^{\epsilon}_{{\rm 0}}(\rho,\mathbf{P}).
\end{aligned}
\end{equation}~$\hfill\blacksquare$

\section{The  POVM-based coherence measures}\label{sec:review}

 For a POVM $\mathbf{E}=\{E_{i}=A_{i}^{\dagger}A_{i}\}_{i=1}^{n}$ on a $d$-dimensional Hilbert space $\mathcal{H}$ \cite{19,20,22}, a canonical Naimark extension projective measurement $\mathbf{P}=\{P_{i}\}_{i=1}^{n}$ of $\mathbf{E}$ is described by a unitary matrix $V$ on Naimark space $\mathcal{H}'$ as
\begin{equation}\label{25}
\begin{aligned}
P_{i}=V^{\dagger}\overline{P_{i}}V,
\end{aligned}
\end{equation}
where
\begin{equation}\label{25}
\begin{aligned}
V=\sum_{i,j=1}^{n}A_{ij}\otimes|i\rangle\langle j|,
\end{aligned}
\end{equation}
with $\{A_{ij}\}_{i,j=1}^{n}$ satisfying the conditions \cite{20}
\begin{equation}\label{25}
\begin{aligned}
\sum_{i=1}^{n}A_{ij}^{\dagger}A_{ik}=\delta_{jk}I_{d},~\sum_{k=1}^{n}A_{ik}^{\dagger}A_{jk}=\delta_{ij}I_{d},~A_{i1}=A_{i},
\end{aligned}
\end{equation}
and
\begin{equation}\label{25}
\begin{aligned}
\overline{P}=\{\overline{P_{i}}=I_{d}\otimes|i\rangle\langle i|\}_{i=1}^{n}.
\end{aligned}
\end{equation}

Let $\mathcal{C}(\rho',\mathbf{\overline{P}})$ be a unitary invariant block-coherence measure, that is,
\begin{equation}\label{8}
\begin{aligned}
\mathcal{C}(\rho',\mathbf{\overline{P}})=\mathcal{C}(U\rho'U^{\dagger},U\mathbf{\overline{P}}U^{\dagger})
\end{aligned}
\end{equation}
for any unitary transformation $U$ on the Hilbert space \cite{19}.
The POVM-based coherence measure $\mathcal{C}(\rho,\mathbf{E})$ of $\rho$ under POVM  $\mathbf{E}$ is defined  \cite{19}
\begin{equation}\label{25}
\begin{aligned}
\mathcal{C}(\rho,\mathbf{E})=\mathcal{C}(\varepsilon(\rho),\mathbf{\overline{P}})=\mathcal{C}(\rho\otimes |1\rangle\langle 1|,\mathbf{{P}}),
\end{aligned}
\end{equation}
where
\begin{equation}\label{8}
\begin{aligned}
\varepsilon(\rho)=\sum_{i,j=1}^{n}A_{i}\rho A_{j}^{\dagger}\otimes|i\rangle\langle j|
\end{aligned}
\end{equation}
is a state on the embedded state Hilbert space $\mathcal{H}_{\varepsilon}$.

From the conclusions in references \cite{19,20,22}, we know that the quantity $\mathcal{C}(\rho,\mathbf{E})$ is a POVM-based coherence measure satisfying the conditions (P1),\ldots,(P4).

Next, we discuss a concrete POVM-based coherence measure.

\begin{proposition}\label{2} Let $\mathbf{E}=\{E_{i}=A_{i}^{\dagger}A_{i}\}_{i=1}^{n}$ be a POVM on the Hilbert space $\mathcal{H}$, the quantity based on the max-relative entropy
\begin{equation}\label{7}
\begin{aligned}
\mathcal{C}_{{\rm max}}(\rho,\mathbf{E})
&=\mathcal{C}_{{\rm max}}(\varepsilon(\rho),\mathbf{\overline{P}})\\
&=\min_{\sigma\in\mathcal{I}_{\rm B}(\mathcal{H}_{\varepsilon})}\log_{2}\min\{\lambda|\varepsilon(\rho)\leq\lambda\sigma\}
\end{aligned}
\end{equation}
is  a block-coherence monotone   and it is quasi-convex. Here $\mathcal{I}_{\rm B}(\mathcal{H}_{\varepsilon})$ is the set of block-incoherent states in the Hilbert space $\mathcal{H}_{\varepsilon}$.
\end{proposition}

\textit{\textbf{Proof.}} We first prove that $\mathcal{C}_{{\rm max}}(\varepsilon(\rho),\mathbf{\overline{P}})$ is  invariant under unitary transformation. The quantity
\begin{equation}\label{7}
\begin{aligned}
\mathcal{C}_{{\rm max}}(\varepsilon(\rho),\mathbf{\overline{P}})=\min_{\sigma\in\mathcal{S}(\mathcal{H}_{\varepsilon})}\log_{2}\min\{\lambda|\varepsilon(\rho)\leq\lambda\sum_{i=1}^{n}\overline{P_{i}}\sigma\overline{P_{i}}\},
\end{aligned}
\end{equation}
where $\sigma$ is an arbitrary  density operator on the  state set $\mathcal{S}(\mathcal{H}_{\varepsilon})$.

For any unitary transformation $U$ on $\mathcal{H}_{\varepsilon}$, we have
\begin{equation}\label{7}
\begin{aligned}
&~~~~\mathcal{C}_{{\rm max}}(U\varepsilon(\rho)U^{\dagger},U\mathbf{\overline{P}}U^{\dagger})\\
&=\min_{\sigma\in\mathcal{S}(\mathcal{H}_{\varepsilon})}\log_{2}\min\{\lambda|U\varepsilon(\rho)U^{\dagger}\leq\lambda U\sum_{i=1}^{n}\overline{P_{i}}U^{\dagger}\sigma U\overline{P_{i}}U^{\dagger}\}\\
&=\min_{\sigma\in\mathcal{S}(\mathcal{H}_{\varepsilon})}\log_{2}\min\{\lambda|\varepsilon(\rho)\leq\lambda\sum_{i=1}^{n}\overline{P_{i}}U^{\dagger}\sigma U\overline{P_{i}}\}\\
&=\min_{\sigma\in\mathcal{S}(\mathcal{H}_{\varepsilon})}\log_{2}\min\{\lambda|\varepsilon(\rho)\leq\lambda\sum_{i=1}^{n}\overline{P_{i}}\sigma\overline{P_{i}}\}\\
&=\mathcal{C}_{{\rm max}}(\varepsilon(\rho),\mathbf{\overline{P}}).
\end{aligned}
\end{equation}

Then, we  show that $\mathcal{C}_{{\rm max}}(\varepsilon(\rho),\mathbf{\overline{P}})$ is a block-coherence monotone.

Firstly we prove that  $\mathcal{C}_{{\rm max}}(\varepsilon(\rho),\mathbf{\overline{P}})\geq 0$, with  equality  if and only if $\varepsilon(\rho)=\sum_{i=1}^{n}\overline{P_{i}}\sigma\overline{P_{i}}$, i.e., $\varepsilon(\rho)$ is the block-incoherent state on $\mathcal{H}_{\varepsilon}$.

By the definition, we known
\begin{equation}\label{7}
\begin{aligned}
\mathcal{C}_{{\rm max}}(\varepsilon(\rho),\mathbf{\overline{P}})=\min_{\sigma\in\mathcal{S}(\mathcal{H})}\log_{2}\min\{\lambda|\varepsilon(\rho)\leq\lambda\sum_{i=1}^{n}\overline{P_{i}}\sigma\overline{P_{i}}\},
\end{aligned}
\end{equation}
since $\varepsilon(\rho)\leq\lambda\sum_{i=1}^{n}\overline{P_{i}}\sigma\overline{P_{i}}$, we have ${\rm Tr}(\lambda\sum_{i=1}^{n}\overline{P_{i}}\sigma\overline{P_{i}}-\varepsilon(\rho))\geq 0$. So $\lambda\geq1$ holds. Hence,
\begin{equation}\label{7}
\begin{aligned}
\mathcal{C}_{{\rm max}}(\varepsilon(\rho),\mathbf{\overline{P}})\geq 0.
\end{aligned}
\end{equation}

According to the properties of maximum relative entropy, the equality holds if and only if $\varepsilon(\rho)=\sum_{i=1}^{n}\overline{P_{i}}\sigma\overline{P_{i}}$, thus $\varepsilon(\rho)$ is a block-incoherent state on $\mathcal{H}_{\varepsilon}$. This implies that $\mathcal{C}_{{\rm max}}(\varepsilon(\rho),\mathbf{\overline{P}})$ satisfies (B1).

The monotonicity of $\mathcal{C}_{{\rm max}}(\varepsilon(\rho),\mathbf{\overline{P}})$ can be easily derived from the properties of the max-relative entropy.
 Hence $\mathcal{C}_{{\rm max}}(\varepsilon(\rho),\mathbf{\overline{P}})$ satisfies (B2).

It is easy to show that $\mathcal{C}_{{\rm max}}(\varepsilon(\rho),\mathbf{\overline{P}})$ is also quasi-convex, i.e.,
\begin{equation}\label{8}
\begin{aligned}
\mathcal{C}_{{\rm max}}(\sum_{i}p_{\varepsilon_{i}}\varepsilon_{i}(\rho),\mathbf{\overline{P}})\leq\max_{i}\mathcal{C}_{{\rm max}}(\varepsilon_{i}(\rho),\mathbf{\overline{P}}),
\end{aligned}
\end{equation}
where $p_{\varepsilon_{i}}={\rm Tr}(K_{i}'\varepsilon(\rho)(K_{i}')^{\dag}),~\varepsilon_{i}(\rho)=\frac{K_{i}'\varepsilon(\rho)(K_{i}')^{\dag}}{p_{\varepsilon_{i}}}$, $\{K_{i}'\}$ is the set of the Kraus operations.

Combining the results above,  we  know that the quantity $\mathcal{C}_{{\rm max}}(\rho,\mathbf{E})$ is a block-coherence monotone and  quasi-convex. ~$\hfill\blacksquare$
\\

Now, we define one-shot block coherence cost under  the maximally POVM-incoherent operations.

\begin{definition} Let $\mathbf{E}=\{E_{i}=A_{i}^{\dagger}A_{i}\}_{i=1}^{n}$ be a POVM on the $d$-dimensional Hilbert space $\mathcal{H}$, and  $\mathbf{P}=\{P_{i}=V^{\dagger}\mathbb{I}\otimes|i\rangle\langle i|V\}_{i=1}^{n}$ be a canonical Naimark extension  of $\mathbf{E}=\{E_{i}\}_{i=1}^{n}$.  We use  $\mathcal{O}$ to denote the set of  the maximally POVM-incoherent operations. For a state $\rho$ and $\epsilon\geq0$, the one-shot block coherence cost under $\mathcal{O}$ is defined as
\begin{equation}\label{6}
\begin{aligned}
\mathcal{C}^{\epsilon}_{\mathcal{O}}(\rho,\mathbf{E})=\min_{\Lambda\in\mathcal{O}}\{\log_{2}N'|F[\Lambda_{\mathcal{O}}(\psi_{N'}),\rho\otimes |1\rangle\langle 1|]\geq 1-\epsilon\},
\end{aligned}
\end{equation}
where $F(\rho,\sigma)=({\rm Tr}[\sqrt{\sqrt{\rho}\sigma\sqrt{\rho}}])^{2}$ is the fidelity between two quantum states $\rho$ and $\sigma$, and the
\begin{equation}\label{6}
\begin{aligned} |\psi_{N'}\rangle=\frac{1}{\sqrt{N'}}\sum_{i=1}^{N'}\frac{P_{i}|\psi_{nd}\rangle}{\sqrt{\langle\psi_{nd}|P_{i}|\psi_{nd}\rangle}} \end{aligned}
\end{equation}
is the maximally block-coherent state,  $|\psi_{nd}\rangle=\frac {1}{\sqrt {nd}}\sum^{nd}_{j=1}|j\rangle$.
\end{definition}

For this  one-shot block coherence cost  under the maximally POVM-incoherent operations the following conclusion holds.

\begin{theorem}\label{1} For quantum state $\rho$ and $\epsilon>0$, we have
\begin{equation}\label{25}
\begin{aligned}
\mathcal{C}^{\epsilon}_{{\rm max}}(\rho,\mathbf{E})\leq\mathcal{C}^{\epsilon}_{\mathcal{O}}(\rho,\mathbf{E})\leq\mathcal{C}^{\epsilon}_{{\rm max}}(\rho,\mathbf{E})+1.
\end{aligned}
\end{equation}
\end{theorem}
\emph{\textbf{Proof.}} Let  $\Delta(\cdot)=\sum_{i}P_{i}\cdot P_{i}$ is the block-dephasing operator of the canonical Naimark extension $\mathbf{P}=\{P_{i}=V^{\dagger}\mathbb{I}\otimes|i\rangle\langle i|V\}_{i=1}^{n}$, with \cite{22}
\begin{equation}\label{25}
\begin{aligned}
V(\rho\otimes|1\rangle\langle1|)V^{\dagger}=\sum_{i,j}A_{i}\rho A_{j}^{\dagger}\otimes|i\rangle\langle j|.
\end{aligned}
\end{equation}

We first prove the left side of Eq.~(82)~. Let $\log_{2}N'=\mathcal{C}^{\epsilon}_{{\mathcal{O}}}(\rho,\mathbf{E})$, and $\sigma=\sum_{i=1}^{n}P_{i}\sigma\otimes|1\rangle\langle1|P_{i}$. The definition of $\mathcal{C}^{\epsilon}_{{\mathcal{O}}}(\rho,\mathbf{E})$ means that there is a maximally POVM-incoherent operation $\Lambda_{{\mathcal{O}}}$ such that $F[\Lambda_{{\mathcal{O}}}(\psi_{N'}),\rho\otimes|1\rangle\langle1|]\geq1-\epsilon$.
Then
\begin{equation}\label{25}
\begin{aligned}
&~~~~\mathcal{C}^{\epsilon}_{{\rm max}}(\rho,\mathbf{E})\\
&=\mathcal{C}^{\epsilon}_{{\rm max}}(\rho\otimes|1\rangle\langle1|,\mathbf{P})\\
&\leq\mathcal{C}_{{\rm max}}(\Lambda_{{\mathcal{O}}}(\psi_{N'}),\mathbf{P})\\
&=\min_{\delta\in\mathcal{S}(\mathcal{H'}),\Delta(\delta)\in\mathcal{I}_{{\rm B}}(\mathcal{H'})}D_{{\rm max}}(\Lambda_{{\mathcal{O}}}(\psi_{N'})\|\Delta(\delta))\\
&=\min_{\sigma\in\mathcal{I}_{{\rm B}}(\mathcal{H'})}D_{{\rm max}}(\Lambda_{{\mathcal{O}}}(\psi_{N'})\|\sigma)\\
&\leq \min_{\sigma\in\mathcal{I}_{{\rm B}}(\mathcal{H'})}D_{{\rm max}}(\Lambda_{{\mathcal{O}}}(\psi_{N'})\|\Lambda_{{\mathcal{O}}}(\sigma))\\
&\leq  \min_{\sigma\in\mathcal{I}_{{\rm B}}(\mathcal{H'})}D_{{\rm max}}(\psi_{N'}\|\sigma)=\log_{2}N'=\mathcal{C}^{\epsilon}_{{\mathcal{O}}}(\rho,\mathbf{E}).
\end{aligned}
\end{equation}
Here $\mathcal{H'}$ is the Naimark space.

Next we prove the right side of Eq.~(82).  Suppose that  the state $\rho'$  satisfies
\begin{equation}\label{25}
\begin{aligned}
\mathcal{C}^{\epsilon}_{{\mathcal{O}}}(\rho,\mathbf{E})
&=\mathcal{C}^{\epsilon}_{{\rm max}}(\rho\otimes|1\rangle\langle1|,\mathbf{P})\\
&=\mathcal{C}_{{\rm max}}(\rho',\mathbf{P})\\
&=\min_{\tau\in\mathcal{I}_{{\rm B}}(\mathcal{H'})}D_{{\rm max}}(\rho'\|\tau)\\
&=\min_{\tau\in\mathcal{I}_{\rm B}(\mathcal{H'})}\log_{2}\min\{\lambda'|\rho'\leq\lambda'\tau\}\\
&=\log_{2}\lambda.
\end{aligned}
\end{equation}
Set $N''=\lceil\lambda\rceil$, then $\rho'\leq N'' \tau$. Consider the following mapping in  the $d$-dimensional Hilbert space $\mathcal{H}$,
\begin{equation}\label{25}
\begin{aligned}
\Lambda(\omega)
&=\frac{1}{N''-1}(N''{\rm Tr}[\psi_{N''}\circ\varepsilon(\omega)]-1)\varepsilon^{-1}(\rho')\\
&~~~+\frac{N''}{N''-1}(1-{\rm Tr}[\psi_{N''}\circ\varepsilon(\omega)])\varepsilon^{-1}(\tau),
\end{aligned}
\end{equation}
where   $\varepsilon(\cdot)$ is  the embedded channel which maps the quantum state from the $d$-dimensional Hilbert space  $\mathcal{H}$ to the $nd$-dimensional Hilbert space. $\varepsilon^{-1}(\cdot)$ is the inverse map of $\varepsilon(\cdot)$. Both $\varepsilon^{-1}(\rho')$ and $\varepsilon^{-1}(\tau)$ are quantum states in the $d$-dimensional Hilbert space $\mathcal{H}$. $\psi_{N''}\circ\varepsilon(\omega)=|\psi_{N''}\rangle\langle\psi_{N''}|\varepsilon(\omega)$, ${\rm Tr}[\psi_{N''}\circ\varepsilon(\omega)]=\langle\psi_{N''}|\varepsilon(\omega)|\psi_{N''}\rangle$.

For all $\omega\in\mathcal{I}_{B}(\mathcal{H})$, we have ${\rm Tr}[\psi_{N''}\circ\varepsilon(\omega)]=\frac{1}{N''}$. The mapping also can be written as
\begin{equation}\label{31}
\begin{aligned}
&~~~~\Lambda(\omega)\\
&=\frac{N''}{N''-1}(1-{\rm Tr}[\psi_{N''}\circ\varepsilon(\omega)])(\varepsilon^{-1}(\tau)-\frac{1}{N''}\varepsilon^{-1}(\rho'))\\
&~~~+{\rm Tr}[\psi_{N''}\circ\varepsilon(\omega)]\varepsilon^{-1}(\rho').
\end{aligned}
\end{equation}

Due to the $\tau\geq\frac{1}{N''}\rho'$, then $\varepsilon^{-1}(\tau)-\frac{1}{N''}\varepsilon^{-1}(\rho')=\varepsilon^{-1}(\tau-\frac{1}{N''}\rho')\geq0$, so $\Lambda$ is  entirely positive. Therefore, there is a positive operator $\Lambda(\omega)$ which is a block-incoherent operation in the Hilbert space $\mathcal{H}$.
Then, we have
\begin{equation}\label{32}
\begin{aligned}
\mathcal{C}^{\epsilon}_{{\mathcal{O}}}(\rho,\mathbf{E})=\log_{2}N''\leq\log_{2}(1+\lambda)\\
\leq\log_{2}\lambda+1=\mathcal{C}^{\epsilon}_{{\rm max}}(\rho,\mathbf{E})+1.
\end{aligned}
\end{equation}~$\hfill\blacksquare$

\section{conclusion}\label{sec:conclusion}

 In the resource theory of block-coherence, we  define a block-coherence measure  $\mathcal{C}_{{\rm max}}(\rho,\mathbf{P})$ based on maximum relative entropy, and  show that it is a coherence monotone and quasi-convex under the maximally block-incoherent operations. The maximally block-coherent state is introduced, and we obtain that the value of $\mathcal{C}_{{\rm max}}(|\psi_{N}\rangle,\mathbf{P})$ only  depends on the number $N$ of projectors in the Hilbert space. Furthermore we give the definition of  the one-shot block coherence cost under the maximally block-incoherent operations and  find the relationship between  the coherence measure $\mathcal{C}_{{\rm max}}(\rho,\mathbf{P})$ and the one-shot block coherence cost.  We describe the deterministic coherence dilution process by constructing block-incoherent  operations based on the resource theory of block-coherence. We also introduce the  coherence measure $\mathcal{C}_{{\rm 0}}(\rho,\mathbf{P})$ based on coherent rank, and obtain the relationship with the one-shot block coherence cost. Based on the  POVM coherence resource theory, we propose a POVM-based coherence measure by using the known scheme of building POVM-based coherence measures from block-coherence measures, and  the one-shot block coherence cost under the maximally POVM-incoherent operations. The relationship between the POVM-based coherence measure and the  one-shot block coherence cost under the maximally POVM-incoherent operations is analysed.

\begin{acknowledgments}
This work was supported by the National Natural Science Foundation of China under Grant No.12071110,
the Hebei Natural Science Foundation of China under Grant No. A2020205014, and Science and Technology Project of Hebei Education Department under Grant Nos. ZD2020167, ZD2021066.
\end{acknowledgments}


\end{document}